\documentstyle[preprint,revtex]{aps}
\begin{document}
\draft
\begin{title}
Possible Spin-Liquid States on the Triangular and Kagom\'{e} Lattices
\end{title}
\author{Kun Yang, L. K. Warman, and S. M. Girvin}
\begin{instit}
Physics Department, Indiana University, Bloomington, IN 47405
\end{instit}
\begin{abstract}
The frustrated quantum spin-one-half Heisenberg model on the triangular and
Kagom\'{e} lattices is mapped onto a single species of fermion carrying
statistical flux $\theta=\pi$. The corresponding Chern-Simons gauge theory is
analyzed at the Gaussian level and found to be massive. This provides a new
motivation for the spin-liquid Kalmeyer-Laughlin wave function. Good overlap
of this wave function with the numerical ground state is found for small
clusters.

\end{abstract}


The 2D spin-${1\over 2}$ antiferromagnetic Heisenberg model has attracted a
lot of interest over the last a few years. It is widely believed that the
ground state has
long-range N\'{e}el order on the square lattice.$^{1}$ On the triangular and
Kagom\'{e} lattices, the situation is much less clear due to the geometric
frustration. In the case of triangular lattice, Huse and Elser have
constructed variational wave functions with long-range order and low energies.
$^{2}$ Numerical evidence was found supporting this scenario.$^{3}$ Recently
however, Singh and Huse$^{4}$ have calculated the sublattice magnetization
using a series expansion method, which is believed to be accurate, and found
that the ground state is nearly disordered for the triangular lattice and
strongly disordered for the Kagom\'{e} lattice due to the large degeneracy of
classical ground state configurations.$^{5}$

Recently a two-dimensional extension of the Jordan-Wigner transformation,
which essentially treats hard core-bosons (see below) as fermions with flux
tubes, or equivalently, fermions coupled to a Chern-Simons gauge field, was
developed.$^{6}$ The advantage of this approach is that the unwanted
hard-core condition in the boson picture is taken care of by the Pauli
principle, but as a price one has to introduce gauge interactions to fix the
statistics. We applied this method to both triangular and Kagom\'{e} lattices.
At mean field level, the flux carried by the fermions is smeared out to form a
uniform background magnetic field. If we neglect the Ising part of the
Hamiltonian (which becomes a nearest-neighbor repulsive interaction between
fermions after the transformation) for the moment, we have noninteracting
fermions moving in a constant magnetic field. Numerical diagonalization yields
two and six Landau subbands, with large excitation gaps$^{23}$ at the Fermi
level of $2\sqrt{3}J$   and $1.46J$   for the triangular and Kagom\'{e}
lattices respectively ($J$ is the coupling between neighboring spins). The
Hall conductance is quantized in such cases, in the form
$\sigma_{xy}={e^{2}\over h}\nu$
where $e$ is the charge carried by the fermions. In the continuum $\nu$ is
just the Landau level filling factor $\nu={\pi\over\theta}$, where $\theta$
is the statistics angle which is $\pi$ in this case. However on a lattice,
$\nu$ is one of the TKKN$^{7}$ integers, not necessarily equal to
${\pi\over\theta}$. If one examines Gaussian fluctuations of the gauge field
about its saddle point (mean field), one can integrate out the fermion
degrees of freedom (which are quadratic in the action) and expand the
effective action for the gauge field about its saddle point to second order to
obtain$^{8}$

\begin{eqnarray*}
S[A_{\mu}]&=&S_{0}+\int{d^{2}xdt[{\epsilon\over 2}E^{2}(\vec{x},t)-{\chi\over
2}B^{2}(\vec{x},t)]}\\
&+& {1\over 4}\int{d^{2}xdt(\nu-{\pi\over\theta})\epsilon_{\mu\nu\lambda}A^
{\mu}F^{\nu\lambda}}+\cdot\cdot\cdot,
\end{eqnarray*}
where $S_{0}$ is the mean field action, $\epsilon$ and $\chi$ are the mean
field values of the long-wavelength, low-frequency dielectric constant and
diamagnetic susceptibility respectively. As noted by Fradkin,$^{8}$ the
fluctuation is massless if and only if the Chern-Simons term in the action is
cancelled, i.e., $\nu={\pi\over\theta}$. This is easy to understand in terms
of self-consistent linear response. Assume there is a long-wave length,
low-frequency fluctuation of the density of the fermions. Since the fermions
carry flux, there should be a fluctuation of magnetic field in the same mode.
According to Maxwell's equations, there will be a nonzero line integral of
electric field around any region $\Gamma$:
\[\oint_{\Gamma}{\vec{E}\cdot d\vec{l}}=-{1\over c}{d\Phi\over dt}=-{h\over e}
{\theta\over \pi}{dN\over dt},\]
where $\Phi$ and $N$ are the flux and number of particles in that region
respectively. Now look at the response of $N$ to this electric field:
\[{dQ\over dt}=e{dN\over dt}=-\oint_{\Gamma}{\sigma_{xy}\vec{E}\cdot d\vec{l}}
=-{\sigma_{xy}\over c}{d\Phi\over dt}.\]
We see the above equations are consistent if and only if $\nu={\pi\over\theta}
$. This is guaranteed in the continuum, where Fetter $et$ $al$ did find a
gapless collective mode.$^{19}$ We have computed the TKKN integer for the
triangular and Kagom\'{e} lattices using the method of MacDonald$^{20}$ and
found in both cases $\nu=-1\not={\pi\over\theta}=1.$
Hence we expect the Chern-Simons field to be massive and the quantum XY model
is therefore likely to have a gap assuming no broken translation symmetry.
This gap may be stable against the Ising perturbation causing the Heisenberg
model to have a spin liquid ground state on these lattices.

By making analogy to the fractional quantum Hall effect (FQHE), Kalmeyer and
Laughlin (KL) suggested a very interesting spin liquid wave function for
the Heisenberg model on the triangular lattice.$^{9,10}$ It was found to have
reasonably good energy (about ten percent higher than the best numerical
estimate). We believe the massiveness of the Chern-Simons theory demonstrated
here provides a more fundamental motivation for quantum Hall physics in
frustrated spin systems. Further more, within the single-mode approximation$^
{12}$ an excitation gap in a 2D system $requires$ a Jastrow-like wave function
whose square is a 2D one-component plasma in order for the structure factor to
vanish at small $q$: $S(q)\sim q^{2}$. SU(2) symmetry uniquely restricts the
coefficient of the plasma charge to be that given by the KL wave function
($m=2$). The spin-${1\over 2}$ excitations argued by Laughlin$^{11}$ similarly
require $m=2$ in the Bose representation. These arguments strongly suggest
that a spin liquid wave function should be of the Kalmeyer-Laughlin type.

In the rest part of this paper we first briefly review the KL wave function,
and prove that it is
equivalent to a projected underlying fermion wave function. We apply this new
wave function to the Kagom\'{e} net, and calculate its energy using the Monte
Carlo method. Then we calculate the overlap between this wave function and the
exact wave
functions on small clusters. Finally we summarize and discuss our results.

The Hamiltonian for antiferromagnetic Heisenberg model is:

\begin{equation}
H=J\sum_{<ij>}{\vec{S}_{i}\cdot\vec{S}_{j}},
\end{equation}
where $J>0$, $\vec{S}_{i}$ is the spin operator at site $i$, and the sum is
over
nearest neighbors. Following Ref[9], we map the spin operators to hard-core
boson operators. The Hamiltonian in this representation is
\begin{eqnarray*}
H&=&T+V\\
T&=&{J\over 2}\sum_{<ij>}{(a_{i}^{\dagger}a_{j}+a_{j}^{\dagger}a_{i})}\\
V&=&J\sum_{<ij>}{n_{i}n_{j}}+{\rm const}
\end{eqnarray*}

Here $N_{s}$ is the number of sites of the lattice, $a_{j}$ is the hard-core
boson operator at $j$th site. As was shown by KL,$^{9}$ on the triangular
lattice, after an
appropriate gauge transformation, this new Hamitonian describes hard core
bosons moving in a uniform magnetic field with field strength one flux
quantum per unit cell, under the symmetric gauge. The bosons  have a
nearest-neighbor repulsive interaction besides the hard core condition. In
the ground state, the number of bosons $N=N_{s}/2$, which means the Landau
level filling factor is one
half. By making analogy to the FHQE, they suggested the trial wave function for
the bosons:

\begin{equation}
\Psi(z_{1},\cdot \cdot \cdot,z_{N})=\prod_{i<j}{(z_{i}-z_{j})^{2}}\prod_{k\le
N}{G(z_{k})e^{-{1\over 4}\vert z_{k}\vert^{2}/l_{0}^{2}}}.
\end{equation}
Here $z_{j}=x_{j}+iy_{j}$ is the complex coordinate of the lattice site
occupied by the $j$th boson, $l_{0}=b(\sqrt{3}/4\pi)^{{1\over 2}}$ is the
magnetic
length which is set to 1 afterwards, $b$ is the lattice constant.
$G(z_{k})=\pm 1$ are gauge phases introduced by the gauge transformation.$^{9}$

The wave function (2) has some nice features, including being a singlet state
in the thermodynamic limit, but it can not be generalized to non-Bravais
lattices, such as the Kagom\'{e} lattice. Also it becomes a singlet only when
the system is infinitely large, so it is not suitable for finite size studies.
For these reasons, we want to find a more general wave function that returns
to (2) on the triangular lattice in the thermodynamic limit, and has better
finite-size properties. To do that, we assume the spins are
carried by spin-${1\over 2}$ fermions that are trapped on lattice sites, and
try to describe state (2) in terms of these underlying fermions.$^{21}$ We can
express the
spin operators in terms of these fermion operators:
\begin{eqnarray*}
a_{j}^{\dagger} & =S_{j}^{+}=c_{j\uparrow}^{\dagger}c_{j\downarrow}\\
a_{j} & =S_{j}^{-}=c_{j\downarrow}^{\dagger}c_{j\uparrow}
\end{eqnarray*}
where $c_{j\uparrow,\downarrow}^{+}$ are creation operators of up(down) spin
fermions at $j$th site. Just as for the bosons, there is a hard core
condition on the fermions: $n_{j\uparrow}+n_{j\downarrow}=1$.

In the second quantized representation, the state (2) is
\[\vert\Psi\rangle =\sum_{\{z_{1},\cdot\cdot\cdot,z_{N}\}}{\Psi(z_{1},\cdot
\cdot\cdot,z_{N})a_{z_{1}}^{\dagger}\cdot\cdot\cdot a_{z_{N}}^{\dagger}\vert
0_{b}\rangle}. \]
Here the sum is over all possible boson configurations. $\vert 0_{b}\rangle$
is the boson vacuum state. Since it corresponds to the state that all spins
are down, we have
\[\vert 0_{b}\rangle =\prod_{j=1}^{N_{s}}{c_{j\downarrow}^{\dagger}}\vert
0_{f}\rangle\]
where $\vert 0_{f}\rangle$ is the fermion vacuum state. So we get
\begin{eqnarray*}
\vert\Psi\rangle &=&{1\over N!}\sum_{\{z_{1},\cdot\cdot\cdot,z_{N};z_{[1]},
\cdot\cdot\cdot,z_{[N]}\}}{\Psi(z_{1},\cdot\cdot\cdot,z_{N})}\\
&\times &c_{z_{1\uparrow}}^{\dagger}c_{z_{1\downarrow}}\cdot\cdot\cdot c_{z_
{N\uparrow}}^{\dagger}c_{z_{N\downarrow}}\prod_{k=1}^{N_{s}}{c_{k\downarrow}
^{\dagger}}\vert 0_{f}\rangle.
\end{eqnarray*}
Here $z_{[j]}=z_{N+j}$ denotes the coordinate of $j$th down spin fermion. The
sum is over all possible fermion configurations satisfying the hard core
condition. We do not distinguish $c_{j}$ and $c_{z}$, $a_{j}$ and $a_{z}$,
$etc$., if $z$ is the complex coordinate of $j$th site. Rearranging the order
of fermion operators and neglecting constant factors, we have
\begin{eqnarray*}
&&\vert\Psi\rangle =\sum_{\{z_{1},\cdot\cdot\cdot,z_{N};z_{[1]},\cdot\cdot
\cdot,z_{[N]}\}}{\Psi(z_{1},\cdot\cdot\cdot,z_{N})}\\
&\times & F(z_{1},\cdot\cdot\cdot,z_{N};z_{[1]},\cdot\cdot\cdot,z_{[N]})\prod_
{k=1}^{N}{c_{z_{k\uparrow}}^{\dagger}}\prod_{l=1}^{N}{c_{z_{l\downarrow}}
^{\dagger}\vert 0_{f}\rangle}.
\end{eqnarray*}
Here $F(z_{1},\cdot\cdot\cdot,z_{[N]})$ is a totally antisymmetric factor:
\begin{eqnarray*}
\vert F(z_{1},\cdot\cdot\cdot,z_{2N})\vert&=&{\rm const}\\
F(\cdot\cdot\cdot,z_{j},\cdot\cdot\cdot,z_{k},\cdot\cdot\cdot,z_{2N}) &=&-F(
\cdot\cdot\cdot,z_{k},\cdot\cdot\cdot,z_{j},\cdot\cdot\cdot).
\end{eqnarray*}
We can take $F$ to be
\[F=\prod_{i<j}{(z_{i}-z_{j})^{-1}(z_{[i]}-z_{[j]})^{-1}}\prod_{k,l\le N}{(z_
{k}-z_{[l]})^{-1}}.\]
If we go back to first quantization, the wave function that describes the
underlying fermions is just

\begin{eqnarray}
&&\Psi_{f}(z_{1},\cdot\cdot\cdot,z_{[N]})=\prod_{i<j\le N}{(z_{i}-z_{j})(z_{
[i]}-z_{[j]})^{-1}}\nonumber\\
&\times & \prod_{k,l\le N}{(z_{k}-z_{[l]})^{-1}}\prod_{m=1}^{N}{G(z_{m})e^{-
{1\over 4}\vert z_{m}\vert^{2}}}.
\end{eqnarray}

Obviously a fermion wave function should be antisymmetric under exchange
(including spin variables), but here we neglect spin variables in the wave
function and treat up and down spin particles as if they were distinguishable,
so the above wave function is legal.$^{18}$ Since we have the hard core
condition on fermions, what we really mean by (3) is the state that is
projected to the subspace with no double occupancy. Hence it is a well
defined wave
function for the original quantum spins.

Now we can use a theorem proved by KL$^{10}$:

\[\prod_{j\ne k}{(\xi_{k}-\xi_{j})}=C_{0}G(\xi_{k})e^{{1\over 4}\vert\xi_{k}
\vert^{2}},\]
where $\xi_{j}$ is the complex coordinate of $j$th site, $C_{0}$ is a
constant,and $G(\xi_{k})$ are the gauge phases.$^{9}$ This theorem holds only
on the triangular lattice in the thermodynamic limit. Using it in (3)
we get up to a constant factor,

\begin{equation}
\Psi_{f}(z_{1},\cdot\cdot\cdot,z_{[N]})=\prod_{i<j\le N}{(z_{i}-z_{j})(z_{[i]}
-z_{[j]})}.
\end{equation}

The right hand side of equation(4) is just the product of two Vandermonde
determinants,$^{21}$ which is what one gets when both spin states of the
first Landau level are fully occupied. Since up and down spin particles occupy
the
same spatial Slater determinant, the resulting state must be a spin singlet,
even after projection.$^{16}$ This provides another way
to prove that the state (2) is a singlet in the thermodynamic limit. The
advantage
of (4) is it gives a singlet even on finite size systems, and it can be
generalized directly to non-Bravais lattices.

As a test of the equivalence between (2) and (4), we calculated the energy of
(4) on a triangular lattice. The calculation of the Ising part of the energy
is straightforward using Monte Carlo, and the total energy is exactly three
times that (for any system size). The extrapolated result is $-0.48\pm 0.01J$
per site, which
agrees with the KL result.$^{9}$ We have found that in our case the data
converges much faster, $i$.$e$., the finite size results are much closer to
the extrapolated result. This tells us that (4) is better for finite size
study. The energy we get for the Kagom\'{e} is $-0.399\pm 0.001 J$ per site,
about eight percent higher than the best numerical estimate.$^{15}$

We have also studied the overlap between (4) and exact ground state on small
clusters, where we need to minimize the finite size effect
by applying periodic boundary conditions (PB). The wave function (4) satisfies
open boundary condition, so we need to solve the wave
functions with PB:

\begin{equation}
t(\vec{L_{j}})\psi(z)=e^{i\phi_{j}}\psi(z);j=1,2,\cdot\cdot\cdot,
\end{equation}
Here $t(\vec{L_{j}})$ is a magnetic translation operator.$^{13}$ This problem
was solved for the torus geometry by Haldane and Rezayi.$^{13}$ It turns out
that the filled Landau level wave function, after neglecting constant factors,
can be expressed in terms of elliptic theta functions$^{13,14}$:

\begin{eqnarray}
&\Psi_{f}&=\prod_{i,j\le N}{[\theta_{1}({\pi\over L_{1}}(z_{i}-z_{j})\vert
\tau)\theta_{1}({\pi\over L_{1}}(z_{[i]}-z_{[j]})\vert\tau)]}\nonumber\\
&\times & \theta_{1}({\pi\over L_{1}}(\sum_{k}{z_{k}-Z_{0}})\vert\tau)\theta
_{1}({\pi\over L_{1}}(\sum_{k}{z_{[k]}-Z_{0}})\vert\tau).
\end{eqnarray}
Here $\theta_{1}(z\vert\tau)$ is the elliptic theta function, $\tau=L_{2}e^
{i\delta}/L_{1}$, $\vec{L_{1}}$ and $\vec{L_{2}}$ are the vectors that
determine the shape of the parallelogram, and $\delta$ is the angle between
them. $Z_{0}$ is the center of mass coordinate which is determined by
$\phi_{1}$ and $\phi_{2}$. In most cases we are interested in, it should be
set to zero.$^{17}$ Like eq. (4), eq. (6) also describes a singlet state. A
truely nondegenerate ground state wave function must be real.$^{9}$ So
instead of using the complex wave function (6) directly, we use $\Psi_{f}e^
{i\phi}+\Psi_{f}^{*}e^{-i\phi}$ in the overlap calculation and use $\phi$ as
a variational parameter. We applied it to several clusters of the triangular
lattice with the shape of a parallelogram (torus geometry) and one with the
shape of a hexagon (twelve spins). The results are listed in table 1. We have
found that the square of the overlap remains large in systems with up to
twenty spins $(4\times 5)$. The energies one gets are close to the Monte
Carlo result, which means the change of boundary condition does not change
the short distance correlations. For reasons we do not understand yet, the
overlap is $exactly$ zero in the $4\times 4$ cluster. We have verified that
$\Psi_{f}$ has the correct symmetries (spin rotation, translation, $180^{0}$
rotation and mirror reflection of space). Apparently there is some additional
hidden symmetry which does not match that of the numerical ground state. We
have done the same calculations on Kagom\'{e} clusters$^{15}$ with twelve and
eighteen spins. Again we got zero overlaps, probably for similar reasons. The
energies, $-0.420J$ and
$-0.418J$ per site for twelve- and eighteen-spin clusters respectively, are
close to the Monte Carlo result.

If the ground state of the triangular lattice
has three sublattice order, such order is suppressed
on clusters $3\times 4$, $4\times 4$ and $4\times 5$ due to incommensurability,
but is not on $6\times 3$ and the hexagon with twelve spins. Our data
suggests this commensurability is a weak effect. Both state (6) and the true
ground states have a lot of symmetries (rotation, translation, etc.). It
could happen that there are so few states of the right symmetries available
that an arbitrary
combination of them will have a decent overlap with the ground state. By
assuming singlet states are
uniformly distributed in the momentum space, we find the number of states
with the right symmetry is of order 1000 in the case of twenty spins, and yet
the square of the overlap is rather large: 0.493. The twelve-spin hexagon has
additional symmetries, so the significance of the remarkably large squared
overlap of 0.966 is unclear.

Quantum Hall wave functions exhibit hidden ODLRO due to the binding of
vortices to charges.$^{24}$ This corresponds to chiral order in the present
problem.$^{25}$ The large overlaps we obtain suggest chiral order is present
in the exact ground states of the clusters.

The central result of this paper is the demonstration that treating spins as
fermions carrying flux tubes leads to a massive Chern-Simons theory on
frustrated lattices. This provides a new and fundamental motivation for
quantum Hall types of spin-liquid physics. We have developed the formalism
needed to compute the
overlap between these wave functions and the exact ground state and we have
obtained significant overlaps for small clusters. It would be highly
desirable to see these calculations extended to large lattices, although this
will require considerable numerical effort.

We acknowledge helpful discussions with Rajiv Singh, V. Elser, A. J.
Berlinsky, C. Kallin and N. Read. This work was supported by NSF-9113911.

\narrowtext
\begin{table}
\caption{overlaps and variational energies on small clusters of the
triangular lattice}
\begin{tabular}{cccc}
cluster &$\vert {\rm overlap}\vert^{2}$ &$E_{v}$ &exact energy \\
\tableline
 hexagon(12) &\dec 0.966 &\dec -0.591 &\dec -0.6103 \\
 $3\times 4$ &\dec 0.821 &\dec -0.519 &\dec -0.5776 \\
 $4\times 4$ &\dec 0.000 &\dec -0.459 &\dec -0.5347 \\
 $6\times 3$ &\dec 0.554 &\dec -0.491 &\dec -0.5811 \\
 $4\times 5$ &\dec 0.493 &\dec -0.481 &\dec -0.5581 \\
\end{tabular}
\label{table I}
\tablenotes{Energies are in unit $J$ per site; $E_{v}$ is the variational
energy.}
\end{table}


\begin{references}
\bibitem{} This can be proved using the Fock cyclic conditions.$^{18}$
\bibitem{} Except for the hexagon case discussed later, where we need a non
zero $Z_{0}$ so that (6) has the right symmetry.
\end{references}
\end{document}